\documentclass[10pt,a4paper]{article}

\usepackage{moreverb,xspace,bm}
\usepackage[left=3cm,right=2cm,top=3cm,bottom=3cm]{geometry}
 \usepackage{tikz}
 \usepackage{rotating}
 \usepackage{float,subfig}
 \usepackage{multicol}
 \usepackage[affil-it]{authblk}
 \usetikzlibrary{arrows,shapes,calc}
 \usetikzlibrary[positioning]
 \usetikzlibrary{patterns}

\newcommand\BibTeX{{\rmfamily B\kern-.05em \textsc{i\kern-.025em b}\kern-.08em
T\kern-.1667em\lower.7ex\hbox{E}\kern-.125emX}}
\newcommand{\ie}{\textit{i.e.\/}\xspace}
\newcommand{\eg}{\textit{e.g.\/}\xspace}

\newcommand{\be}{\begin{eqnarray}}
\newcommand{\ee}{\end{eqnarray}}

\newcommand{\bea}{\begin{eqnarray*}}
\newcommand{\eea}{\end{eqnarray*}}

\newcommand{\ben}{\begin{enumerate}}
\newcommand{\een}{\end{enumerate}}

\newcommand{\bref}[1]{(\ref{#1})}
\newcommand{\breq}[1]{\bref{eqn:#1}}

\newcommand{\bns}{\begin{normalsize}}
\newcommand{\ens}{\end{normalsize}}
\newcommand{\bfs}{\begin{footnotesize}}
\newcommand{\efs}{\end{footnotesize}}

\begin{document}

\newcommand\Tspace{\rule{0pt}{2.6ex}}
\newcommand\Bspace{\rule[-1.2ex]{0pt}{0pt}}

\newcommand{\cip}{\mbox{$\perp\!\!\!\perp$}}
\newcommand{\nip}{\not\!\!\!\,\cip}
\newcommand{\todo}[1]{{\sc***#1***}\\}

\title{Bayesian regression discontinuity designs: Incorporating clinical knowledge in the causal analysis of primary care data} 

\author[1]{Sara Geneletti%
\thanks{E-mail address: \texttt{s.geneletti@lse.ac.uk}; Corresponding author}}
\author[2]{Aidan G.~O'Keeffe}
\author[3]{Linda D.~Sharples}
\author[4]{Sylvia Richardson}
\author[2]{Gianluca Baio}

\affil[1]{\small Department of Statistics, London School of Economics and Political Science, WC2A 2AE, UK}
\affil[2]{\small Department of Statistical Science, University College London, WC1E 6BT, UK}
\affil[3]{\small Leeds Institute of Clinical Trials Research, University of Leeds, LS2 9JT, UK}
\affil[4]{\small MRC Biostatistics Unit, Cambridge, CB2 0SR, UK}

\maketitle

\begin{abstract}
  The regression discontinuity (RD) design is a quasi-experimental
  design that estimates the causal effects of a treatment by
  exploiting naturally occurring treatment rules. It can be applied in
  any context where a particular treatment or intervention is
  administered according to a pre-specified rule linked to a
  continuous variable. Such thresholds are common in primary care drug
  prescription where the RD design can be used to estimate the causal
  effect of medication in the general population. Such results can
  then be contrasted to those obtained from randomised controlled
  trials (RCTs) and inform prescription policy and guidelines based on
  a more realistic and less expensive context. In this paper we focus
  on statins, a class of cholesterol-lowering drugs, however, the
  methodology can be applied to many other drugs provided these are
  prescribed in accordance to pre-determined guidelines. Current NHS
  guidelines state that statins should be prescribed to patients with
  10 year cardiovascular disease risk scores in excess of twenty
  percent. If we consider patients whose risk scores are close to the
  twenty percent risk score threshold we find that there is an element
  of random variation in both the risk score itself and its
  measurement. We can therefore consider the threshold a randomising
  device that assigns statin prescription to individuals just above
  the threshold and withholds it from those just below. Thus we are
  effectively replicating the conditions of an RCT in the area around
  the threshold, removing or at least mitigating confounding. We frame
  the RD design in the language of conditional independence which
  clarifies the assumptions necessary to apply an RD design to data,
  and which makes the links with instrumental variables clear. We also
  have context specific knowledge about the expected sizes of the
  effects of statin prescription and are thus able to incorporate this
  into Bayesian models by formulating informative priors on our causal
  parameters.

\end{abstract}

\maketitle

\footnotetext[2]{Email: \texttt{s.geneletti@lse.ac.uk}}

\section{Introduction}

The regression discontinuity (RD) design is a quasi-experimental
design that estimates the causal effects of a treatment by exploiting
naturally occurring treatment rules. Since its inception in the 1960's
in educational economics \cite{ThistlethwaiteCampbell1960}, the RD
design has successfully been applied in areas of economics, politics,
and criminology
\cite{vanderKlaauw2002,vanderKlaauw2008,Lee2008,BerkdeLeeuw1999}
amongst others. More recently, it has been reworked in the econometric
causal inference literature \cite{ImbensLemieux2008,HahnToddVDK2001}
and there has been some interest in the design in epidemiology
\cite{Rutter2009, Finkelsteinetal1996} and health economics \cite{HEC:HEC3027}.

The RD design can be applied in any context where a particular
treatment or intervention is administered according to a pre-specified
rule linked to a continuous variable --- referred to as the assignment
variable. Such thresholds are common in many fields and, in
particular, in primary care drug prescription. For instance, according to the National Institute for Clinical Excellence (NICE) guidelines \cite{NICE}, in the UK statins (a class of cholesterol-lowering drugs) should be prescribed to patients with 10 year cardiovascular disease (CVD) risk scores in excess of
twenty percent. Consider
patients whose risk scores are close to the twenty percent risk score
threshold; typically there is an element of random variation in both
the risk score itself and its measurement. Thus, we can consider the
threshold to be a randomising device that assigns treatment --- statin
prescription --- to individuals just above the threshold and withholds
treatment from those just below the threshold. In other words, if we
focus on an area close to the threshold then we have a situation
that is analogous to a randomised controlled trial, resulting in removal or
mitigation of confounding where we can identify and estimate causal
effects of treatments in primary care. 

The RD design can be useful in situations where evidence from randomised controlled
trials (RCTs) is available, as it is often the case that RCT results
are not consistently replicated in primary care. In such situations the RD design can shed a light on
why this might be the case. In other contexts, RD designs can confirm RCT results where other
observational data might have failed to do so. Furthermore, RD methods,
whilst not providing as substantive evidence of a causal effect as an
RCT, are cheaper to implement, can be typically applied to much larger
datasets and are not subject to as many ethical constraints. This
could make such methods desirable in the overall accumulation of
evidence regarding the effectiveness of a particular treatment,
administered using strict prescription guidelines, on an outcome of
interest in primary care. Finally, there are many situations where RCTs cannot be run, for example, in the
case of experimental treatments for terminal diseases. The RD design
means that doctors can administer the treatments to the severely ill but
still obtain a valid (if local) causal effect of the treatment,
provided they adhere to a strict guideline.

In this paper our focus is two-fold. Firstly we formulate the RD design
in the framework of conditional independence. This has, as yet, not
been done and we believe that it both clarifies the underlying
assumptions and makes explicit the link with instrumental variables
(IVs) of which the RD design is a special case.

Secondly, we introduce a Bayesian analysis of the RD design based on
the prescription of statins in primary care. While Bayesian methods
have been applied to the RD design, work has been principally on
spline models \cite{Koo:97, Holmes:01}. We focus here on models incorporating prior information
which have not been widely considered, especially in primary care
contexts. Since much is known already about the effect of statins on
LDL cholesterol, due principally to RCTs, we believe that this example is
a good starting point for the application of the Bayesian methods as
strong prior information on the effect of statins is
available. Furthermore, as part of the analysis we are interested in
estimating a causal effect for GPs who adhere to guidelines. This
requires us to think carefully about formulating priors that are
informative of the process that drives adherence. While the existence
of robust information in this context facilitates the formulation of
prior models, this is by no means a pre-requisite of this
methodology. We note that our principal motivation is not to replicate
the results of RCTs or to solely estimate the causal effect of statins
on LDL cholesterol using an RD design. Rather, we are interested in
considering Bayesian methodology in an RD design and use the effect of
statin prescription on LDL cholesterol as a motivating example.

We consider three models, each of which is informative to a different
degree, and examine how sensitive the results are to prior
specification in datasets of different sizes. The discussion of the
results highlights the importance of thinking carefully about prior
specification and also that, in some contexts, it is not difficult to
formulate reasonable prior beliefs. 

We use simulated data based closely on actual statin prescriptions in
the THIN primary care database to showcase our Bayesian
methodology. The simulation algorithm we develop allowed us to set the
size of the treatment effect and control the levels of unobserved
confounding and the strength of the RD design. However we were also
able to retained the structure and idiosyncrasies of the original data
by adding our modifications to these data rather than simulating it in
its entirety. 

The paper is organised in three parts: in the first one, section \ref{sec:rd101}, we first describe the RD design in more detail and introduce the running example (statins prescription for the primary care prevention of cardiovascular disease). Then, we formalise the assumptions necessary to identify a causal treatment effect using the RD design. Finally, we clarify the links between the RD design and instrumental variables and introduce the causal estimators. 

The second part of the paper (section \ref{sec:bayes}) introduces the details of our novel Bayesian model formulation. In this section, we describe and justify all the distributional assumptions used in our model and discuss the implications of incorporating prior clinical knowledge in causal analyses, specifically when they are based on the RD design. 

Finally, in the third part of the paper (sections \ref{sec:sims} and \ref{sec:results}) we introduce a simulation study, providing details of the simulation strategy as well as the results of the application of our models to the simulated data. Problems and extensions are discussed in section \ref{sec:discussion}.

\section{The regression discontinuity design}
\label{sec:rd101}
\subsection{The basics of the RD design}

In its original inception, the RD design was used to evaluate the effect of schooling on a number of adult life outcomes \eg income. The classic example considers scholarships that are offered to students according to their grade-point-average or other markers of academic/sporting ability. However, the RD design can be applied in any context where an intervention, be it a drug, a lifestyle modification, or other, is administered according to guidelines based on continuous variables. 

These situations also arise commonly in primary care drug
prescription: examples include the prescription of anti-hypertensive
drugs when blood pressure exceeds 140/90mmHg1 or of selective
serotonin reuptake inhibitors for patients exhibiting more than 4
symptoms in the ICD-10 classification of depression. Another
interesting case, which we use as a running-example in this paper, is
the prescription of statins, a class of cholesterol-lowering drugs, in
the primary prevention of cardiovascular disease, in the UK. There are clear NICE
guidelines regarding statin prescription \cite{NICE}, which makes this
a suitable case-study to show the potential of the RD design to
perform a causal analysis using primary care data. In the case of statins,
the guidelines recommend that individuals who have not experienced a
cardiac event should be treated if their risk of developing
cardiovascular diseases (CVD) in the subsequent 10 years, as
predicted by an appropriate risk calculator (\eg Framingham risk
calculator), exceeds 20\%. Note that in the original NICE guideline, the choice of the threshold was driven also by cost-effectiveness considerations.

A 10 year cardiovascular risk
score is predicted based on a logistic regression with a number of clinical
and lifestyle factors. These typically include, amongst others,
blood pressure, total cholesterol, smoking status. Thus the RD design can be used to estimate
the effect of statins on clinical outcomes, specifically LDL cholesterol levels, in individuals around this threshold level.

\subsubsection{The sharp RD design}

\begin{figure}[!h]
  \centering
  \includegraphics[width=14cm]{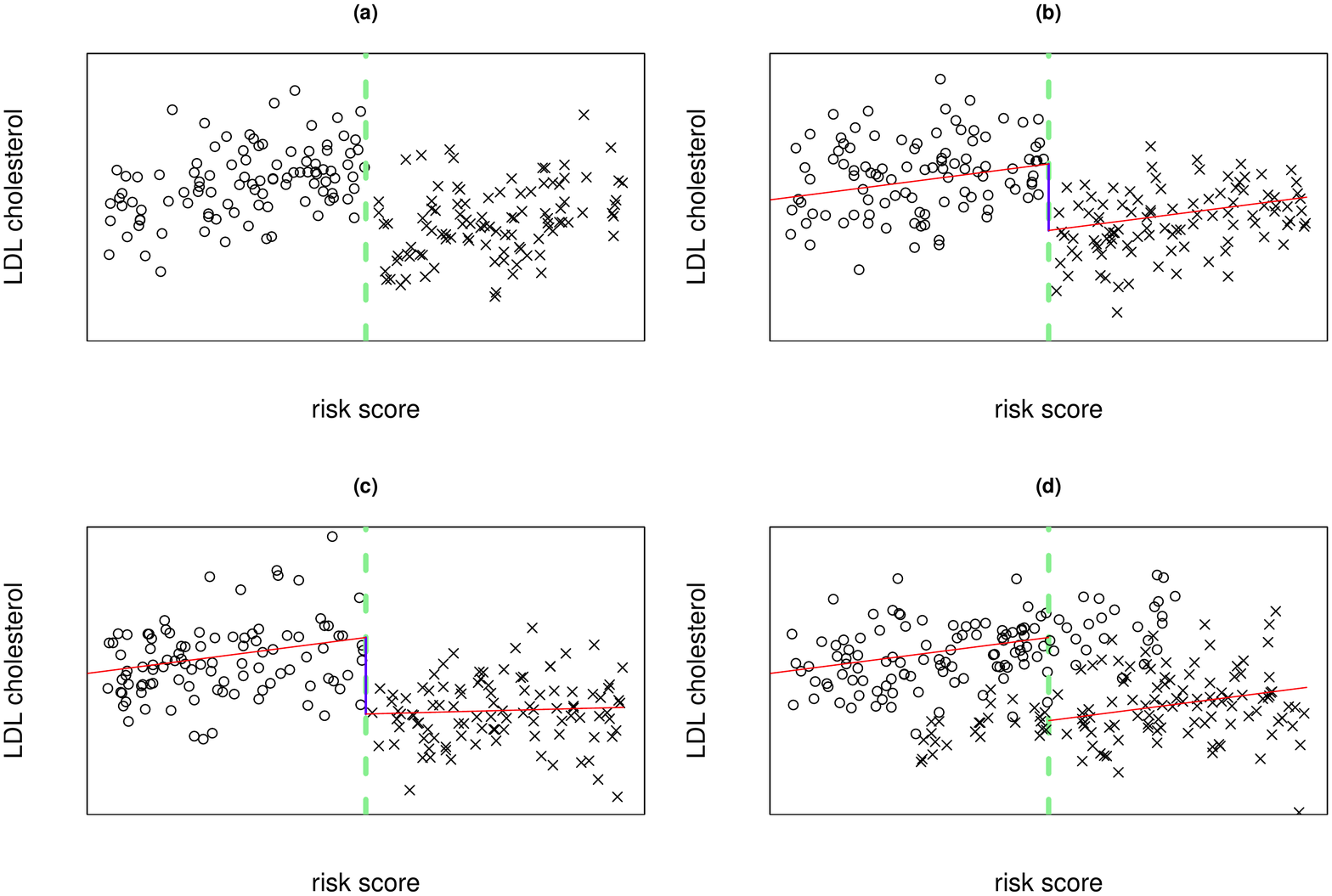}
\caption{(a) The sharp RD design with crosses indicating patients who have been prescribed statins and circles those who have not, (b) The sharp RD design with equal slopes with regression lines above and below the threshold and a bold vertical bar at the threshold to indicate the effect size, (c) The sharp design with different slopes, (d) The fuzzy design. Note that there are crosses below and circles above the threshold indicating that some GPs are not adhering to the treatment guidelines.  \label{fig:rd1}}
\end{figure}

In an ideal situation, all general practitioners (GPs, UK family
doctors) prescribe statins to patients who have a risk score above the
20\% risk score threshold and do not prescribe the drugs to those
whose risk score falls below 20\%. In addition, if statins
also have a positive effect (\ie they reduce LDL cholesterol) then a plot
of risk score versus LDL cholesterol could look like Figure
\ref{fig:rd1}(a), particularly if cholesterol is linear in the risk
score. Here, circles represent untreated patients and crosses treated
patients.
The ``jump'' at the 20\% risk score can then be interpreted as the
average treatment effect at the threshold. If we assume that the
effect of statins is constant across risk scores i.e.~that the slope of the regression of LDL cholesterol against risk score is the same above and below the threshold then the effect at the threshold can be
considered an average effect for all risk scores as in Figure \ref{fig:rd1}(b).
It is possible however, that the slopes differ depending on whether the patient is above or below the threshold. In this case, the scatter plot of LDL cholesterol against risk score might look like Figure \ref{fig:rd1}(c). 

In this situation, where thresholds are strictly adhered to the RD design is
termed \textit{sharp} and the value of the jump is estimated and interpreted as the causal effect.

\subsubsection{The fuzzy RD design}

Typically in most applications, and particularly in the case of statin prescription, the RD design is not sharp. This is because GPs will often prescribe statins to patients below the threshold if they deem that it will be beneficial or possibly not prescribe statins to patients above the threshold if statins are counter-indicated for these patients. We term this GP \textit{adherence} to the guidelines. We contrast this to the situation where patients are not \textit{complying} to treatment prescribed. We make this distinction in order to avoid confusion by using the term compliance to describe the GP's behaviour when typically this term is used to describe patients' behaviour. For the remainder of the paper and, in particular, for the simulations, we assume that patients comply to their prescription. In Section \ref{sec:discussion} we briefly highlight differences between these two types of compliance and discuss how we might account for patient non-compliance in the real data. When GPs do not adhere to treatment guidelines, a plot of risk score against cholesterol might look like Figure \ref{fig:rd1}(d) where the crosses below the threshold and circles above the threshold indicate individuals who are not being prescribed according to the guidelines. In this situation the RD design is termed \textit{fuzzy}. In order to estimate treatment effects (typically local/complier effects) additional assumptions must be made as detailed in Section \ref{sec:assumptions} below.

\subsection{Assumptions}
\label{sec:assumptions}

A number of assumptions must hold in order for the RD design to lead to the identification of causal effects. These assumptions are expressed in different ways depending on the discipline \cite{ImbensLemieux2008,Leeetal2004,HahnToddVDK2001}. We describe our approach in the language of conditional independence \cite{APD.03,APD.79,Didelezetal2010}: in our view, this approach helps to clarify situations where the RD design can be used and highlights the links with the theory of instrumental variables. Throughout the paper, we follow standard notation: if a variable $A$ is independent of another $B$ conditional on a third $C$, then $p(A,B\mid C)=p(A\mid C)p(B\mid C)$ and we write $A \cip B\mid C$ \cite{APD.79}. 

Let $X$ be the assignment variable on which the treatment guidelines are based. Specifically, if $x_0$ is the threshold given by the treatment guidelines, then let $Z$ be the threshold indicator such that $Z=1$ if $X>x_0$ and $Z=0$ if $X \leq x_0$. Furthermore, let $T$ indicate the treatment administered (prescribed); we assume a binary treatment, so that $T=1$ means treatment is administered (prescribed) and $T=0$ means it is not. Also, let $\bm{C}=\{\bm{O}\cup\bm{U}\}$ be the set of confounders, where $\bm{O}$ and $\bm{U}$ indicate fully observed and partially or fully unobserved variables, respectively. Finally, $Y$ is the continuous outcome variable. In our case study, $X$ is the 10 year cardiovascular risk score with $x_0=0.2$. Thus $Z=1$ if a patient's 10 year risk score exceeds 0.2 and $0$ if their score is below 0.2. The treatment is statin prescription (NB: \textit{not} patient taking the treatment). The outcome of interest is the level of LDL cholesterol. 

We discuss in detail the assumptions necessary for the RD design, below.
\begin{enumerate}
\item[A1.] \textit{Association of treatment with threshold indicator}:\\
Treatment assignment must be associated with the treatment guidelines. This assumption can be expressed equivalently as
\[Z \nip T,\]
which implies that $Z$ and $T$ are not marginally independent. In the context of statin prescription this assumption will hold if the GPs adhere to the NICE treatment guidelines and $Z$ is predictive of treatment $T$, \ie they prescribe the treatment to patients with a 10 year risk score that exceeds 20\% and do not prescribe statins to patients whose risk score is below 20\%. This assumption can be tested directly by estimating the association between $Z$~and~$T$. This does not mean that the RD design breaks down when GPs prescribe according to their own criteria, as the guideline itself is still in place. What happens if some GPs prescribe according to their own criteria is that assumption A1 becomes weaker as the association between the threshold indicator (\ie the guideline) and prescription practice decreases. However, provided the association is still strong, \ie a sufficient number of GPs adhere to it, fuzzy methods can be brought to bear.  

\item[A2.] \textit{Independence of guidelines}:\\
The treatment guidelines cannot depend on any of the characteristics of the patient (excluding $X$), \ie they cannot be changed for individual patients. We can express this assumption in terms of the threshold indicator as
\[Z \cip \bm{C},\]
\ie $Z$ is marginally independent of $\bm{C}$ --- and we note that this should hold at least around the threshold. We can also see this assumption as meaning that the patient characteristics (excluding $X$) cannot determine their value of $Z$. 

Assumption A2 does not preclude dynamic treatment strategies as long as these are pre-specified. We could consider a dynamic strategy as one that depends not only on the risk score but on a number of factors. For instance, a GP might look at a fixed number of (observed and recorded) risk factors when deciding whether to prescribe statins and only prescribe when a pre-specified minimum number indicate elevated risk. This will be different for each patient but will not be different for two patients with the same values for the risk factors.

If the threshold indicator is associated with some unobserved confounders $\bm{U}$, a weaker version of this assumption is that the threshold indicator does not depend on the unobserved confounders given the observed confounders $\bm{O}$
\[Z \cip \bm{U} \mid \bm{O}.\] 
We can think of this as the RD design applied within strata of the observed confounders, for example by considering statin prescription for men only. 

Neither version of A2 can be tested as each involves either implicitly or explicitly the unobserved confounders $\bm{U}$. However, A2 is likely to hold in one of the two forms, because it is typically externally imposed and does not vary from patient to patient or from one GP to another.

\item[A3.] \textit{Unconfoundedness}:\\
In order for the RD design to be a valid randomisation device, the outcome must be independent of the threshold indicator, conditionally on the other variables. This can be expressed more formally as
\be
\label{eqn:ass3}
Y \cip Z\mid (T,\bm{C}).
\ee
For the statin example, this requires that patients cannot determine their treatment assignment, \ie that even when they know about the treatment rule, they cannot manipulate their outcome in order to fall above or below the treatment threshold. This guarantees that there is some randomness in where subjects fall with respect to the threshold. While it is plausible for patients to try and persuade their GPs to prescribe statins when they do not have a high enough risk score, this is unlikely to happen in a systematic manner and can also be subsumed in a weakening of assumption A1. Nevertheless, (\ref{eqn:ass3}) breaks down if the GPs systematically fail to adhere to the risk score guideline but rather base treatment decisions on unobserved factors. As total cholesterol is part of the risk factor and LDL cholesterol is in turn a part of the total cholesterol, it might appear that assumption A3 does not hold. However, total cholesterol also includes HDL cholesterol and the risk score contains a number of other factors. Thus the link between $Y$ and $Z$ in our example is not deterministic but subject to random variation generally and most importantly for individuals around the threshold, $\bm{C}$ will contain all the remaining confounders such as HDL cholesterol and thus there will be no direct link.

The condition in (\ref{eqn:ass3}) is also untestable as it too implicitly involves the unobserved confounders $\bm{U}$. It is therefore important to consider whether individuals on either side of the threshold really can be considered to be exchangeable.

\item[A4.] \textit{Continuity}:\\
It is necessary to assume that, conditionally on the other variables, the expected outcome is continuous around the threshold $x_0$. This can be expressed in terms of
\[\mbox{E}(Y\mid Z,X=x,T,\bm{C}) \mbox{ is continuous in $x$ (at $x_0$) for $T =0,1$}.\]
To understand why this assumption is necessary note that the marginal expectation of the outcome, conditionally on the assignment variable alone, \ie $\mbox{E}(Y\mid X=x)$, is in fact \textit{discontinuous} around the threshold and it is the size of the discontinuity that is interpreted as a causal effect. The continuity of the conditional expectation guarantees that it is the threshold indicator and not any of the other variables that is responsible for the discontinuity in the outcome. Some RD design texts \cite{HahnToddVDK2001}, state this assumption in terms of the the limits from above and below of the expectation of $Y$. More generally, we can assume that the conditional distribution of $Y$ given the two treatments (active and control) and the assignment is continuous at the threshold \cite{ImbensLemieux2008}. This assumption is partly testable on the observed confounders $\bm{O}$, \eg if partial regression plots of the outcome against observed confounders conditional on the assignment exhibit discontinuities around the threshold, then assumption A4 is called into question.

In the context of statin prescription, this assumption requires that the expected value of LDL cholesterol as a function of variables other than the risk be continuous. If there were a discontinuity in the association between LDL cholesterol and, for instance, body mass index (BMI) conditionally on the risk score being 20\%, then it would not be possible to attribute the jump in LDL cholesterol to the threshold indicator and as a consequence the treatment. In particular, if BMI is a confounder for the relationship between risk score and LDL cholesterol, it would follow that the discontinuity observed in LDL cholesterol could be due to BMI.

\item[A5.] \textit{Monotonicity} (fuzzy design only):\\
For the fuzzy design, another assumption is necessary in order to identify a \textit{local} causal effect rather than an \textit{average} effect (we formally define these in Section \ref{sec:iv} below). This assumption requires that there are no GPs who systematically prescribe the opposite of what the guidelines recommend. We define the pair of prescription strategies that a GP has prior to seeing a patient as $(S_a,S_b)$, for $a$bove and $b$elow the threshold, respectively. These are binary decision variables taking value 1 if the GP prescribes the treatment and 0 otherwise. Then we can express the monotonicity assumption as
\[\Pr(S_a=0,S_b=1)=0,\]
\ie the probability of there being GPs who would decide to prescribe the treatment to \textit{all} individuals below the threshold and who would decide not to prescribe the treatment to individuals above the threshold is 0. We must also assume that the GPs act according to these prescription strategies. In the potential responses literature this is often referred to as the ``no defiers'' assumption. There are a number of weaker versions of the monotonicity assumption (see, for example, \cite{deChaisemartin2012,SmallTan2007})  which are plausible in some RD design settings when the strong assumption given above cannot be assumed to hold. 

In the context of our running example, this seems a very plausible assumption: even if a GP is not in agreement with the guidelines, he or she will be concerned with patient benefit rather than in compliance with NICE recommendations. However, if we allow for patient non-compliance to the treatment, then the monotonicity assumption implies that there are no patients who on principle will decide to do the opposite of what they are prescribed. It is likely that there are some of these patients in a real context and thus the weaker assumptions can be invoked. We discuss these briefly in Section \ref{sec:discussion}. It is not generally possible to test this assumption unless we are able to inquire of GPs or patients how their decision strategy is formulated. 
\end{enumerate}

\subsection{Links with IV and causal effect estimators}
\label{sec:iv}

It is well-known that the RD design threshold indicator $Z$ is a
special case of a \textit{binary instrumental variable}
\cite{ImbensLemieux2008,Didelezetal2010}.
We link the RD design to the IV framework using the language of
conditional independence and thereby clarify how the RD design fits
into the context of experiments and quasi-experiments.

Consider the case of a binary treatment (\eg an active drug treatment versus a placebo) and the two experimental designs commonly used for causal inference. The first is the ``gold standard'', the double blinded randomised controlled trial with perfect compliance (henceforth RCT), meaning that the individuals in the trial take exactly and only the treatment to which they have been assigned. The second is the randomised controlled Trial but with Partial Compliance (henceforth TPC), meaning that not all the individuals take the treatment they have been assigned. 

In the RCT it is possible to estimate the \textit{average treatment (causal) effect} 
\begin{eqnarray}
\mbox{ATE} &=& \mbox{E}(Y\mid T=1)-\mbox{E}(Y\mid T=0)
\label{eqn:ate} \nonumber\\ 
&=& \mbox{E}(Y\mid Z=1)-\mbox{E}(Y\mid Z=0), 
\label{eqn:atez}
\end{eqnarray}
without making additional assumptions, since randomisation and perfect compliance guarantee (bar unlucky and unlikely lack of balancing) that any difference in the outcome is due only to the treatment assigned. 

However, in the TPC an average causal effect cannot be estimated because there is confounding by treatment self-administration. This means that some patients in the treatment arm (and we typically do not know which ones) have not actually taken the treatment or, conversely (and often less likely), that some of the patients in the control arm have obtained the treatment and taken it. We will not know what motivated the patients to act as they did and thus an intention-to-treat effect will typically be a biased estimate of the average treatment effect. 

In these situations, a \textit{local} (sometimes called a \textit{complier}) \textit{average treatment effect}
\begin{eqnarray}
\label{eqn:late}
\mbox{LATE}= \frac{\mbox{E}(Y\mid Z=1)-\mbox{E}(Y\mid Z=0)}{\mbox{E}(T\mid Z=1)-\mbox{E}(T\mid Z=0)}.
\end{eqnarray}
is estimated using the treatment assignment as an IV. In order to make
the LATE identifiable, it is necessary that the RD monotonicity
assumption A5~holds.

By comparing the RD design to the RCT and TPC scenarios described
above, we see that the RD sharp design is analogous to the RCT and
that the fuzzy RD design is analogous to the TPC with the treatment
assignment corresponding to the threshold indicator. Thus, in a sharp
RD the ATE is equivalent to (\ref{eqn:atez}), while for the case of
the fuzzy design, where the threshold guidelines are not always
adhered to, the LATE is a measure of the threshold effect with the
threshold indicator as an IV. 

This correspondence highlights the appropriateness of the ATE and
LATE as causal effect estimates in the primary care context. The ATE
is clearly the appropriate causal estimate for the sharp design as
this is equivalent to the RCT. For the fuzzy design, the ATE as shown
in \breq{atez} corresponds to the the intention-to-treat (ITT)
estimator in a TPC. This ITT estimator is subject to confounding and
does not identify a causal effect. Therefore the LATE is typically
identified in TPCs as and is thus appropriate for the fuzzy RD. 

In our context, the LATE identifies the causal effect for those patients registered with GPs whose prescription strategy corresponds with NICE guidelines. We
have no reason to believe that the types of patients registered with such GPs are systematically different to the patients of GPs whose
strategies are different. Thus we believe that the LATE provides us
with a valid and potentially generalisable causal effect estimate. A
further discussion, involving lack of patient compliance to treatment
is given in Section~\ref{sec:discussion}.

\section{Bayesian model specification}
\label{sec:bayes}
Our motivation for using Bayesian methods to analyse data generated in
a RD setting, is three-fold. Firstly, the Bayesian framework enables
us to set priors in such a way as to reflect our beliefs about the
values of the parameters and potentially impose substantively
meaningful constraints on their values. For example, given extensive
RCT literature \cite{HTA}, it is widely accepted that the
effect of statin treatment is a decrease in LDL cholesterol of about 2
mmol/l. When modelling the LATE, we can parameterise the numerator
(\ie the sharp treatment effect ATE) in such a way as to express this
belief, while still allowing for uncertainty around this informed
prior guess. We discuss strategies for achieving this goal in Section
\ref{sec:modelsATE}.

A second reason for adopting a Bayesian approach is that, when estimating the LATE, a major concern is that the denominator, that is, the difference between the probabilities of treatment above and below the threshold, can be very small at the threshold (\ie when the threshold  is a weak instrument). The Bayesian framework allows us to place prior distributions on the relevant parameters in such a way that the difference must exceed a certain minimum or such that the difference is ``encouraged'' to exceed a certain minimum. This can stabilise the LATE estimate, as we discuss in Section~\ref{sec:modelsDenom}. 

Finally, standard frequentist methods rely on complex asymptotic arguments to estimate the variance associated with the treatment effect, which often results in overly conservative interval estimations. By contrast, Bayesian analyses are typically implemented using MCMC methods, which allow increased flexibility on the modelling structure, as well as relatively straightforward estimation for all the relevant quantities (either directly specified as the parameters of the model, or derived using deterministic relationships among them).

The inclusion of (relatively) strong prior information makes sense especially in contexts where the signal in the data is particularly weak and confounded and when, as in the RD design context, information about both the drug treatment and the probability of treatment above and below the threshold is available through previous research and extensive content-matter knowledge. We discuss the strength of effect of the prior information when looking at the results of the analysis and the simulation studies, as well as to what extent results from these studies can be considered reliable in Section \ref{sec:sims}. 

As the results appear to be more sensitive to priors on the denominator of the LATE, we summarise the priors for the ATE briefly in Section \ref{sec:modelsATE} before tackling the prior models on the denominator in more detail in \ref{sec:modelsDenom}. 

\subsection{Local linear regression}
\label{sec:llr}
The estimators we consider depend on linearity assumptions which do not always hold for the whole range of the threshold variable. This can put too much weight on data far from the threshold, thereby resulting in biased estimates. In this case, we can either consider more flexible estimators such as splines or we can explore local linear regression (LLR) estimators.

LLR estimates are obtained using data only within some fixed bandwidth, $h$, either side of the threshold. This achieves three aims: \textit{(i)} to use data around the threshold so that points further away have little or no influence on the predictions at the threshold; \textit{(ii)} to make linearity assumptions more plausible, as a smaller range of points is used, which belong to an area where linearity is more likely to hold; \textit{(iii)} to obtain smooth estimates. There are some recommendations in the literature regarding the optimal size of a bandwidth \cite{ImbensLemieux2008}, however these appear somewhat arbitrary.

\subsection{Models for the ATE}
\label{sec:modelsATE}
In line with equation \bref{eqn:ate}, we estimate the average LDL cholesterol level as a function of the threshold indicator. First, we model the observed LDL cholesterol level separately for the individuals below (whom we indicate with $l=b$) and above ($l=a$) the threshold, as 
\[ y_{il} \sim \mbox{Normal}(\mu_{il}, \sigma^2) \]
and specify a regression on the means
\begin{eqnarray} 
\label{eqn:sharpest}
\mu_{il} = \beta_{0l} + \beta_{1l}x^{c}_{il},
\end{eqnarray}
where $x^{c}_{il}$ is the centered distance from the threshold $x_0$ for the $i-$th individual in group $l$. 

Obviously, the observed value of $x^{c}_{il}$ determines whether, under perfect GP adherence, the individual is given the treatment or not. Thus, for $l=a,b$, the expressions in \bref{eqn:sharpest} are equivalent to $\mbox{E}(Y\mid Z=1)$ and $\mbox{E}(Y\mid Z=0)$ respectively, and the ATE may be written 
\begin{eqnarray}
\label{eqn:ATE}
\mbox{ATE} = \Delta_\beta =: \beta_{0a} - \beta_{0b},
\end{eqnarray} 
that is the difference in the two averages \textit{at} the threshold, \ie when $x^{c}_{il}=0$.

Within the Bayesian approach, to complete the model specification we also need to assign suitable prior distributions to the parameters $(\beta_{0l},\beta_{1l},\sigma^2)$. Where possible, we use the information at our disposal to assign the values of the priors for the model parameters. For example, we know the plausible ranges of the risk score and the LDL cholesterol. We also know from previous studies, trials and conversations with clinicians, that LDL cholesterol increases with risk score and that once statins are taken, the LDL cholesterol tends to decrease. We attempt to encode this information in the priors below.

With (at least moderately) large datasets, the posterior inference is less sensitive to the distributional assumptions selected for the variance $\sigma^2$, because there is enough information from the observed data to inform its posterior distribution. As a result, we consider a relatively vague uniform prior on the standard deviation scale: $\sigma \sim \mbox{Uniform}(0,5)$. 

We consider the following specification for the coefficients for the regression models above and below the threshold:
\begin{eqnarray}
\label{eqn:prior1}
\beta_{0b}  \sim  \mbox{Normal}(m_0,s_0^2) \qquad & \mbox{and} & \qquad \beta_{1b} \sim \mbox{Normal}(m_{1b},s_{1b}^2) \\
\label{eqn:prior2}
\beta_{0a}  = \beta_{0b}+ \phi \mbox{\hspace*{1.2cm}}\qquad & \mbox{and} & \qquad \beta_{1a} \sim \mbox{Normal}(m_{1a},s_{1a}^2). 
\end{eqnarray}
The priors on the parameters $\beta_{0b}$ and $\beta_{1l}$ for $l \in \{a,b\}$ are chosen so such that they result in LDL cholesterol levels that are with plausible values for the observed range of risk scores. This can be achieved by selecting suitable values for the hyper-parameters $(m_0,m_{1b},m_{1a},s^2_0,s^2_{1b},s^2_{1a})$\footnote{For instance, selecting $m_0=3.7$, $m_{1b}=8$, $s_0=0.5$ and $s_{1b}=0.75$ implies that the prior 95\% credible interval for the estimated LDL level ranges in $[2.57;4.83]$mmol/l for individuals with a risk score of 0 and in $[2.72;4.68]$mmol/l for individuals close to the threshold. For the slope $\beta_{1a}$ above the threshold, we encode the assumption that the treatment effect is subject to a form of ``plateau'', whereby for individuals with very high risk score, the effect is marginally lower than for those closer to the threshold. See the online supplementary material for details.}.

The parameter $\phi$ represents the difference between the intercepts at the threshold, \ie ``jump'' due to the causal effect of the treatment. We consider two different specifications for $\phi$ upon varying the levels of informativeness on the prior distribution 
\begin{eqnarray*}
\phi^{wip}\sim\mbox{Normal}(0,2) \qquad \mbox{and}  \qquad \phi^{sip}\sim\mbox{Normal}(-2,1)
\end{eqnarray*}
The former assumes that on average, the treatment effect is null as the magnitude of the prior variance is in this case large enough that the data can overwhelm the null expectation and thus we identify it as weakly informative prior (\textit{wip}). We indicate with the notation $\Delta_\beta^{\rm{\it{wip}}}$ the ATE estimator expressed in the form of equation \bref{eqn:ATE} resulting from this formulation of the priors. 

In the latter, we encode information coming from previously observed evidence that statins tend to have an effect of around 2 mmol/l at the threshold. In this particular case study, given the extensive body of RCTs on the effectiveness of statins, we set the variance to 1, which essentially implies relatively strong belief in this hypothesis. We term this the strongly informative prior (\textit{sip}) and the resulting ATE estimator is $\Delta_\beta^{\rm{\it{sip}}}$. 

\subsection{Models for the denominator of the LATE}
\label{sec:modelsDenom}
Since we know that in clinical practice there is a clear possibility that the assignment to treatment does not strictly follow the guidelines, as there may be other factors affecting GPs decisions, we also construct a suitable model to compute the LATE estimator. To do so, we need to estimate the denominator of equation \bref{eqn:late}. We start by considering the total number of subjects treated on either side of the threshold, which we model for $l \in \{a,b\}$ as
\[ \sum_{i=1}^{n_l} t_{il} \sim \mbox{Binomial}(n_l,\pi_l),  \]
where $n_l$ is the sample size in either group. The quantities $\pi_a$ and $\pi_b$ represent $\mbox{E}(T\mid Z=1)$ and $\mbox{E}(T\mid Z=0)$, respectively and thus can be used to estimate the denominator of equation \bref{eqn:late} as 
\begin{eqnarray}
\label{eqn:Denom}
\Delta_\pi=:\pi_a-\pi_b.
\end{eqnarray}
As we have little information, a priori, on the probabilities of prescription above and below the threshold, we consider three different prior specifications for the parameters $\pi_l$, leading to three possible versions of the denominator~$\Delta_\pi$. We investigate the sensitivity of results to different beliefs regarding the strength of the threshold instrument by acting on the difference $\Delta_\pi$ directly.

\subsubsection{Unconstrained prior for $(\pi_a,\pi_b)$}
First, we consider a simple structure, in which the probabilities on either side of the threshold are modelled using vague and independent prior specifications. For convenience, we use conjugate Beta distributions that are spread over the entire $[0;1]$ range
\[\pi_l \sim \mbox{Beta}(1,1).\]
Since this specification does not impose any real restriction on the estimation of the probabilities $\pi_l$, we term this model unconstrained (\textit{unc}) and we indicate the denominator resulting from the application of \bref{eqn:Denom} under this prior as $\Delta^{\rm{\it{unc}}}_\pi$.

\subsubsection{Fixed difference prior for $(\pi_a,\pi_b)$}
Next, we show how to build priors that effectively impose a sharp design on the data. This means that $\pi_a \neq \pi_b$ and the LATE estimator does not ``explode''. 

We consider a structure that implies a fixed distance between $\pi_a$ and $\pi_b$ --- consequently we termed this the fixed difference prior (\textit{fix}). The idea is to use informative and correlated conjugate Beta distributions linking $\pi_a$ and $\pi_b$, specifically in the form 
\begin{eqnarray*}
\pi_b & \sim & \mbox{Beta}(\alpha_b,[n_b+1]) \\
\pi_a & \sim & \mbox{Beta}(\alpha_a,1)	\\
\alpha_b & \sim & \mbox{Uniform}(1,100000) \\
\alpha_a & = & \nu + \alpha_b \\
\nu &\sim & \mbox{Uniform}(200,10000). 
\end{eqnarray*}
Since the parameters of the Beta distribution can be thought of as representing the prior number of ``successes''$-1$ and ``failures''$-1$ respectively, the prior for $\pi_b$ encodes the (extreme and simplistic) assumption that all $n_b$ subjects below the threshold are untreated, \ie all of the observed untreated. Similarly, the prior for $\pi_a$ implies that there are no untreated above the threshold. In addition, we are also imposing at least 200 more treated above than below, by modelling the parameter $\nu$ uniformly in the range $[200;10000]$. As a consequence, we imply that the resulting denominator (which we indicate as $\Delta^{\rm{\it{fix}}}_\pi$, under this formulation) has a minimum value which depends on the size of the bandwidth through the sample size~$n_b$. 

This model is, in fact, quite strict as it imposes in the prior a situation that is very similar to a sharp RD design and thus it would not be a very sensible choice in a ``real'' analysis --- but we consider it here to investigate the effect of an extreme prior specification on the results.

\subsubsection{Flexible difference prior for $(\pi_a,\pi_b)$}
\begin{figure}[!h]
\begin{center}
\includegraphics[scale=0.5]{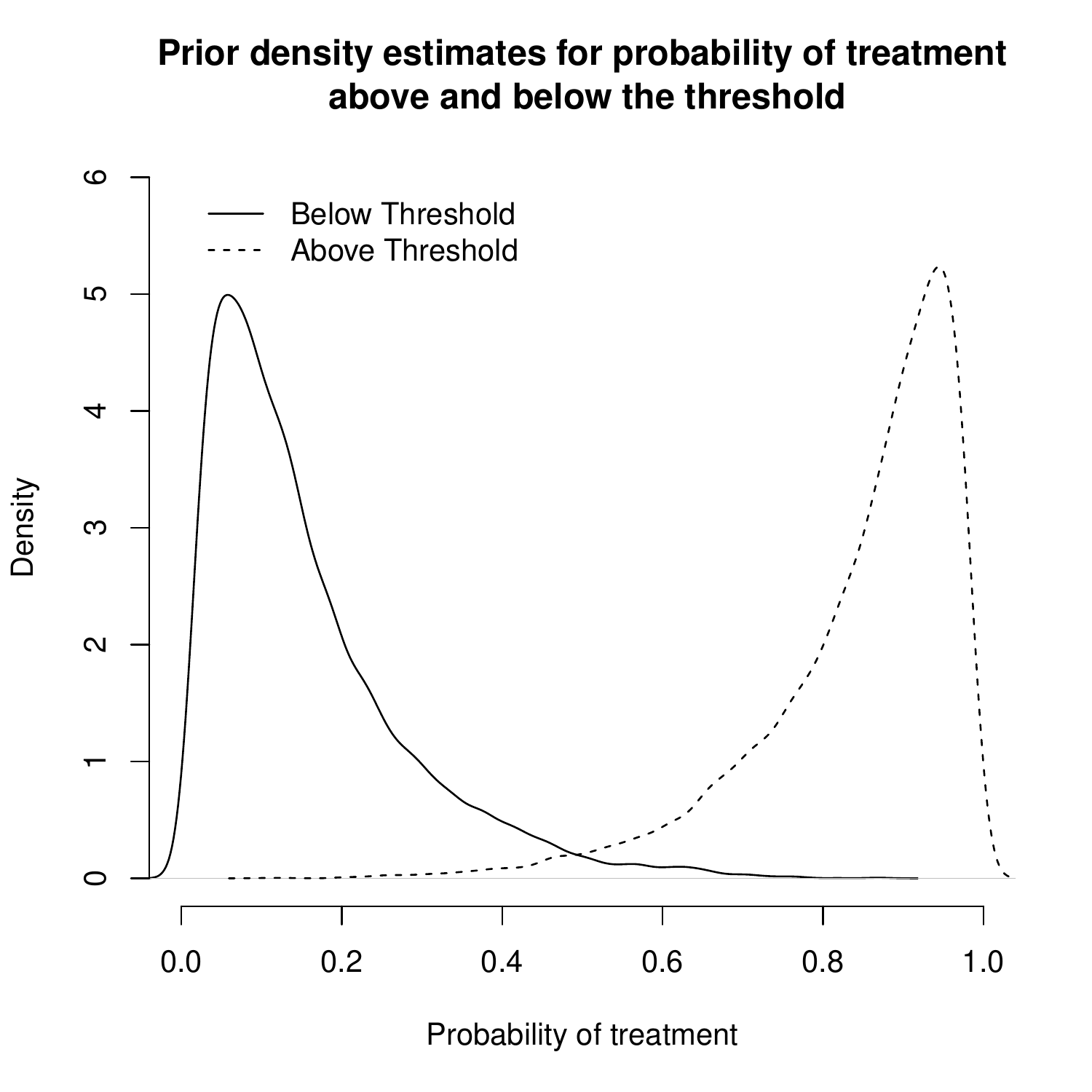}
\end{center}
\caption{Prior predictive distribution for the probability of treatment below (solid line) and above (dashed line) induced by the flexible difference model. The former is substantially lower than the cut-off value of 0.5, while the latter mostly exceeds this. Nevertheless, both allow for the full range of values in $[0;1]$ to be possible \label{fig:prior3}}\label{prior3}
\end{figure}

Finally, we construct a model in which prior information is used in order to ``encourage'' a significant difference between the probabilities --- we term this the flexible difference prior (\textit{fdp}) and define it as
\begin{eqnarray*}
\mbox{logit}(\pi_a) \sim \mbox{Normal}(2,1) \qquad \mbox{and} \qquad \mbox{logit}(\pi_b) \sim \mbox{Normal}(-2,1)
\end{eqnarray*}
These priors imply that, effectively, we assume the probability of treatment below the threshold to be substantially lower than 0.5 (\ie most of the probability mass is concentrated in the range $[0,0.5]$, while still allowing for a small chance of exceeding this interval). Similarly, we assume that most of the probability mass for $\pi_a$ is above the cut-off value of 0.5, as shown in Figure \ref{fig:prior3}. In this way, we limit the possibility that the two quantities are the same, a priori, while not fixing a set difference between them. The denominator derived using these prior assumptions is indicated by $\Delta^{\rm{\it{fdp}}}_\pi$. 

\subsection{Models for the LATE}
It is easy to obtain suitable estimates for the LATE by simply combining the models of \S\ref{sec:modelsATE} and \S\ref{sec:modelsDenom}. We tried a number of combinations of different specifications. Eventually, we chose three as they were representative of the results. In all cases, the numeerator is given by $\Delta^{\rm{\it{sip}}}$ as the results were not sensitive to changes in the ATE. We combined
\begin{itemize}
\item the fixed difference model in the denominator with the strongly informative prior model for the ATE and termed this the \textit{constrained} model
\[\mbox{LATE}_{\rm{\it{cnst}}} = \frac{\Delta^{\rm{\it{sip}}}_\beta}{\Delta^{\rm{\it{fix}}}_\pi}; \]
\item the flexible difference model in the denominator with the strongly informative prior in the numerator and term this the \textit{flexible} model
\[\mbox{LATE}_{\rm{\it{flex}}} = \frac{\Delta^{\rm{\it{sip}}}_\beta}{\Delta^{\rm{\it{fdp}}}_\pi}; \]
\item the unconstrained denominator with the strongly informative prior in the numerator and term this the \textit{unconstrained} model 
\[\mbox{LATE}_{\rm{\it{unct}}} = \frac{\Delta^{\rm{\it{sip}}}_\beta}{\Delta^{\rm{\it{unc}}}_\pi}. \]
\end{itemize}


\section{Simulated data}
\label{sec:sims}


We consider the simulation of data for which an RD design would be
appropriate. We are interested in testing our methodology on data that
are as close as possible to the data on primary care prescriptions.
One reason is that results based on realistic data, with all its
idiosyncrasies and quirks, are potentially of more value than
simulations based on pre-specified regression models. Another reason
is that these data retain the basic structure of the original data so
that the ranges of the variables of interest, LDL cholesterol levels,
risk scores etc. are, for the most part, within the true levels of these
variables. This means that it makes sense to think about prior
information for the simulated data in much the same way as one would
for the real data as it is as noisy as the real data and retains its quirks. See Jones et
al. \cite{Jonesetal2012} for examples of Bayesian methods for weak IVs
which use data simulated from the ground up.

Specifically we base our simulation scheme on \textit{The Health Improvement Network} (THIN) dataset (see www.thin-uk.com). The THIN database is one of the largest sources of primary care data in the United Kingdom and consists of routine, anonymised, patient data collected at 497 GP practices. Broadly, the dataset is representative of the general UK population and contains patient demographics and health records together with prescription and therapy records, recorded longitudinally.
Our aim is to use the models presented in Section \ref{sec:bayes} to estimate a pre-defined treatment effect of the prescription of statins on LDL cholesterol level (mmol/L). We base our simulation scheme on a subset of data from THIN consisting of males only aged over 50 years ($N = 5720$ records). It is important to consider the assumptions presented in Section \ref{sec:assumptions} whilst ensuring that the characteristics of the simulated data reflect the original observations. With this in mind, our simulation method can be split broadly into two parts. 

\begin{figure}[!h]
\begin{center}
\tikzstyle{block} = [rectangle, draw, fill=white!20, 
    text width=30em, 
text centered, 
 rounded corners, minimum height=2em]
\tikzstyle{line} = [draw, -latex']
    
\begin{tikzpicture}[node distance = 2.1cm, auto]
    \node [block] (step1) {\textit{Part I, 1}. Select real data sample and define variables to be used.};
    \node [block, below of=step1] (step2) {\textit{Part I, 2}. Remove any existing observed treatment effect in the data sample and obtain  $y^{\rm{SIM 1}}$. \\ {\small $y^{\rm{SIM 1}}$ has the same expectation as $y$ the observed LDL cholesterol but without association with treatment or threshold} };
    \node [block, below of=step2] (step3) {\textit{Part I, 3}. Fit a glm for P(Treatment) using the real data. };
    \node [block, below of=step3] (step4) {\textit{Part I, 4}. Use this glm to estimate P(Treatment), $\hat{p}$,\\  {\small adjusting model parameters pertaining to HDL cholesterol level and threshold to obtain the desired level of unobserved confounding and threshold instrument strength}.\\ Hence draw an estimated treatment indicator for each individual, $\hat{t}_i$.};
		\node [block, below of=step4] (step5) {\textit{Part II, 1}. Regress LDL cholesterol level $y^{\rm{SIM 1}}$ on estimated treatment indicator, keeping the estimated residuals from the fitted model: $\hat{\varepsilon}_{2i}$};
		\node [block, below of=step5] (step6) {\textit{Part II, 2}. Regress residuals, $\hat{\varepsilon}_{2i}$, from the above model on risk score, age and a diabetes indicator. Use the fitted values from this model to distort the LDL cholesterol values to obtain $y^{\rm{SIM 2}}$};
		\node [block, below of=step6] (step7) {\textit{Part II, 3}. Add a treatment effect on LDL cholesterol level $y^{\rm{SIM 2}}$ with desired expected value $\tau$ when $\hat{t}_i  = 1$ to obtain $y^{\rm{SIM 3}}$ the final simulated value of LDL cholesterol};

    \path [line] (step1) -- (step2);
		\path [line] (step2) -- (step3);
		\path [line] (step3) -- (step4);
		\path [line] (step4) -- (step5);
		\path [line] (step5) -- (step6);
		\path [line] (step6) -- (step7);
\end{tikzpicture}
\caption{A flow-diagram describing the main steps in the simulation algorithm. \label{fig:flow}}
\end{center}
\end{figure}
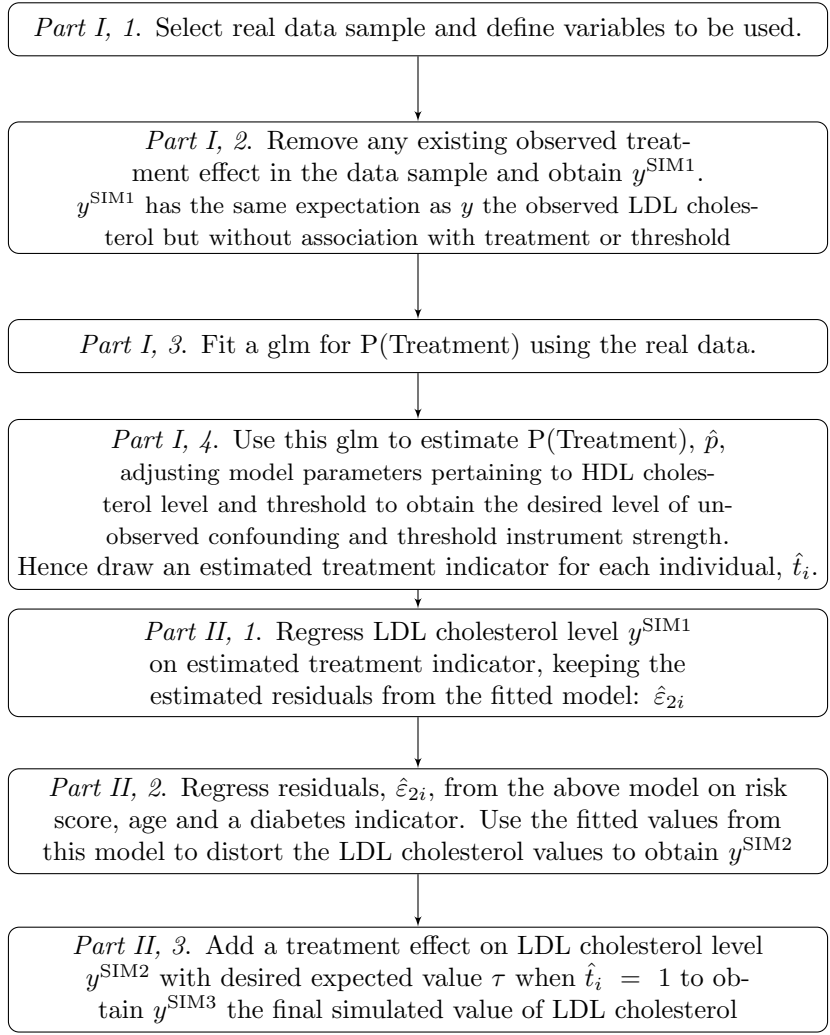

The first part concerns the adjustment of the original subset from THIN to remove any pre-existing treatment effect and the simulation of a treatment indicator. This is done in an effort to ensure that any differences between these groups are due mainly to the assignment of treatment to one group during the simulation algorithm. The second part concerns the simulation of the outcome (LDL cholesterol level). As previously described in Section \ref{sec:assumptions}, we use a threshold of a 0.2 10-year cardiovascular risk score. We also include extra randomness at various points in the simulation algorithm, which is described in Sections \ref{sec:sim1} and \ref{sec:sim2} below. A flow-diagram describing the main steps is displayed in Figure \ref{fig:flow}.

We note here that our simulation method is carefully constructed and somewhat different from that seen in many simulation studies. This is done so that the simulated data retains much of the variability present in the original data upon which the simulations are based. One of our main aims is to consider RD designs that could be fitted to real-life observational data and so we feel it is important that the simulated data reflect this.

\subsection{Simulation algorithm: Part I (simulating the treatment)}
\label{sec:sim1}
\begin{enumerate}
\item From the available subsample of THIN, we consider, for the $i-$th individual, the following observed variables: $x^c_{i}$ as the centered version of the 10-year cardiovascular risk score; $z_{i}$ as the threshold indicator, such that $z_{i} = 1$ if $x^c_{i} > 0$ (\ie if the uncentered risk score $x_i>0.2$) and $z_{i} = 0$ otherwise; $t_{i}$ as the treatment indicator, where $t_{i} = 1$ if the individual receives statins and $t_{i} = 0$ otherwise; $a_{i}$ as the observed age; $d_i$ as an indicator of diabetes, taking value 1 if the $i-$th individual is diabetic and 0 otherwise; $h_i$ as the individual HDL cholesterol level (measured in mmol/l); and $y_{i}$ as the individual's LDL cholesterol level. These are the starting point for our simulations.

\item We then remove any pre-existing effects of treatment and threshold on the outcome $y_{i}$. This will allow the comparison of estimated results to a known, pre-specified, treatment effect (which we describe in Part II) as well as justifying the assumption that the outcome does not depend on the threshold. We fit the normal linear model:
 \begin{equation}\label{nlm1}
   y_{i} = \alpha_{0}+\alpha_{1}t_{i}+\alpha_{2}z_{i}+\varepsilon_{1i}
 \end{equation}
with the $\varepsilon_{1i}$ ($i \in \{1,\ldots,N\}$) independent, normal, zero mean error terms with a constant variance. We define $\bm{y}$ to be the vector of observed LDL cholesterol values, with sample mean $\bar{y}$, and $\hat{\bm{y}}$ as the vector of fitted values from the model (\ref{nlm1}). We draw a new set of simulated LDL cholesterol outcome values, denoted $\bm{y}^{\rm{SIM1}}$ such that, for $i = 1,\ldots,N$:
\[
 y_{i}^{\rm{SIM1}} = y_{i} - \hat{y}_{i} + w_{i}
\]
with $w_{i}$ drawn at random from a $\mbox{Normal}(\bar{y},0.1^{2})$ distribution. We add the $w_{i}$ term so that $y_{i}^{\rm{SIM1}}$ has approximately the same expectation as $Y_{i}$ with the variance term included to reflect additional uncertainty in $y_{i}^{\rm{SIM1}}$ but still reasonably small because we do not expect $y_{i}^{\rm{SIM1}}$ to differ too greatly from  $y_{i}$. A normal distribution is assumed for LDL cholesterol values, in general, throughout.

\item At this point, we define the individual probability of treatment $p_{i} = \Pr(T_{i} = 1)$ and fit the following generalised linear model:
\[
\log\left(\frac{p_{i}}{1-p_{i}}\right) = \alpha_{3}+\alpha_{4}a_{i}+\alpha_{5}d_{i}+\alpha_{6}x^{c}_{i}+\alpha_{7}h_{i}+\alpha_{8}z_{i}.
\]
Let $\hat{\bm{\alpha}}$ denote the estimates from the fitted model with corresponding estimated covariance matrix $\Sigma_{\bm\alpha}$. We re-draw the parameter estimates from a $\mbox{Normal}(\hat{\bm{\alpha}}, \Sigma_{\bm\alpha})$ distribution, which we indicate as $\tilde{{\bm\alpha}}$. This is usually done when imputing data to reflect our uncertainty in the estimated values of $\bm{\alpha}$. We replace $\tilde{\alpha}_{7}$ and $\tilde{\alpha}_{8}$ with pre-specified values to adjust the level of confounding and the strength of threshold as an instrument for treatment, respectively, in the simulated dataset. 

\item For each individual, we estimate $p_{i}$ (denoting the estimate $\hat{p}_{i}$) and randomly draw an estimated treatment variable, $\hat{t}_{i}$ from a $\mbox{Bernoulli}(\hat{p}_{i})$ distribution. 
\end{enumerate}

\subsection{Simulation algorithm: Part II (simulating the outcome)}
\label{sec:sim2}
\begin{enumerate}
\item From Part I, we form the following normal linear model for the adjusted LDL cholesterol values
\[
y^{\rm{SIM1}}_{i} = \gamma_{0}+\gamma_{1}\hat{t}_{i}+\varepsilon_{2i},
\]
with the $\varepsilon_{2i}$ ($i \in \{1,\ldots,N\}$) independent normal, zero mean, error terms with a constant variance. We also obtain the vector of estimated residuals, $\hat{\bm{\varepsilon}}_{2}$. 

\item We fit the normal linear model:
\[
\hat{\varepsilon}_{2i} = \gamma_{2} + \gamma_{3} a_{i}+\gamma_{4}d_{i}+\gamma_{5}x^{c}_{i}+\varepsilon_{3i}
\]
with the $\varepsilon_{3i}$ ($i \in \{1,\ldots,N\}$)  independent, normal, zero mean error terms with a constant variance. We calculate the vector of fitted values from this model and add to each fitted value its corresponding standard error estimate, denoting the resulting vector $\tilde{\bm{\varepsilon}}_{2}$. This is done to slightly perturb the fitted values of $\bm{\varepsilon}_{2}$, incorporating additional randomness and uncertainty into the estimated values. We then add this $\bm{\varepsilon}_{2}$ to $\bm{y}^{\rm{SIM1}}$ to form a slightly distorted vector of simulated LDL cholesterol values, which we denote as $\bm{y}^{\rm{SIM2}}$.

\item Finally, we add a treatment effect of a pre-specified size, $\tau$ by defining:
\[
y^{\rm{SIM3}}_{i} = y_{i}^{\rm{SIM2}} + (1-\hat{t}_{i})v_{1i}+\hat{t}_{i}v_{2i},
\]
where $v_{1i} \sim \mbox{Normal}(0,0.5^{2})$ and $v_{2i} \sim \mbox{Normal}(\tau,0.5^{2})$. The resulting vector $\bm{y}^{\rm{SIM3}}$ is a set of simulated LDL cholesterol values with a treatment effect of size $\tau$ for the treated. A relatively small variance of 0.25 is chosen here so that the treatment effect can be distinguished.
\end{enumerate}

\subsection{Unobserved Confounding}
We aim to examine the properties of the estimators presented in Section \ref{sec:bayes} under varying levels of unobserved confounding. We use the HDL cholesterol level as an unobserved confounder because it is predictive of both LDL cholesterol and treatment. There are, of course, many other variables that could  have been included in the regression models we used to formulate the simulated outcomes, but for the sake of simplicity we only considered one confounder at this stage.

The estimated correlation between the LDL and HDL cholesterol levels is 0.18 in the original dataset and, to increase unobserved confounding, we also use an adjusted dataset as a basis for simulation, where the estimated correlation between the LDL and HDL cholesterol levels is augmented to 0.5. In addition, the level of unobserved confounding is adjusted in Step 3 of Part I of the simulation algorithm, when the value of $\tilde\alpha_{7}$ is selected. Given these two ways of manipulating the relationship between HDL cholesterol, treatment probability and LDL cholesterol, we define four levels of unobserved confounding, where unobserved confounding increases with level number. The levels are  determined by the values of $r =$ Corr(HDL, LDL) in the simulated data and $\tilde{\alpha}_{7}$ in Step 3 of the simulation algorithm (Part I) and we list them according to $(r, \tilde{\alpha}_{7})$ as: Confounding Level 1 (0.18, 4), Confounding Level 2 (0.5, 4), Confounding Level 3 (0.18, -2) and Confounding Level 4 (0.5, -2). 

The scenarios identifying confounding level 3 and 4 above correlate higher HDL cholesterol with lower probability of being treated, whilst maintaining the positive correlation between HDL and LDL cholesterol. Thus, the effect of the confounder acts in opposite directions on the probability of treatment and the size of the outcome. This is not entirely realistic in the context of statin prescription; nevertheless, we are interested in assessing the performance of the estimators even in such an extreme scenario. Simple plots of the  risk score against LDL cholesterol level that identify treated and untreated patients, such as those in Figures \ref{fig1} and \ref{fig2} (Column 1), are useful in assessing the level of compliance of a GPs within a particular dataset.

\subsection{Threshold as an Instrumental Variable}
In our analysis, we consider cases in which the threshold acts as either a \textit{strong} or a \textit{weak} instrumental variable for the treatment. This is achieved through the pre-defined choice of $\tilde\alpha_{8}$ in Step 3 of Part I of the simulation algorithm. In particular, $\tilde\alpha_{8} = 10$ and $\tilde\alpha_{8} = 4$ are chosen to imply a strong and a weak instrument, respectively. When $\tilde{\alpha}_{8} = 10$, the probability of being treated is almost certain above the threshold, however, when $\tilde\alpha_{8} = 4$ the effect of the instrument is of the same order as that of the confounders, and the probability of being treated is much smaller.

\section{Results}
\label{sec:results}
\begin{figure}[!h]
\begin{center}
\includegraphics[scale=0.7]{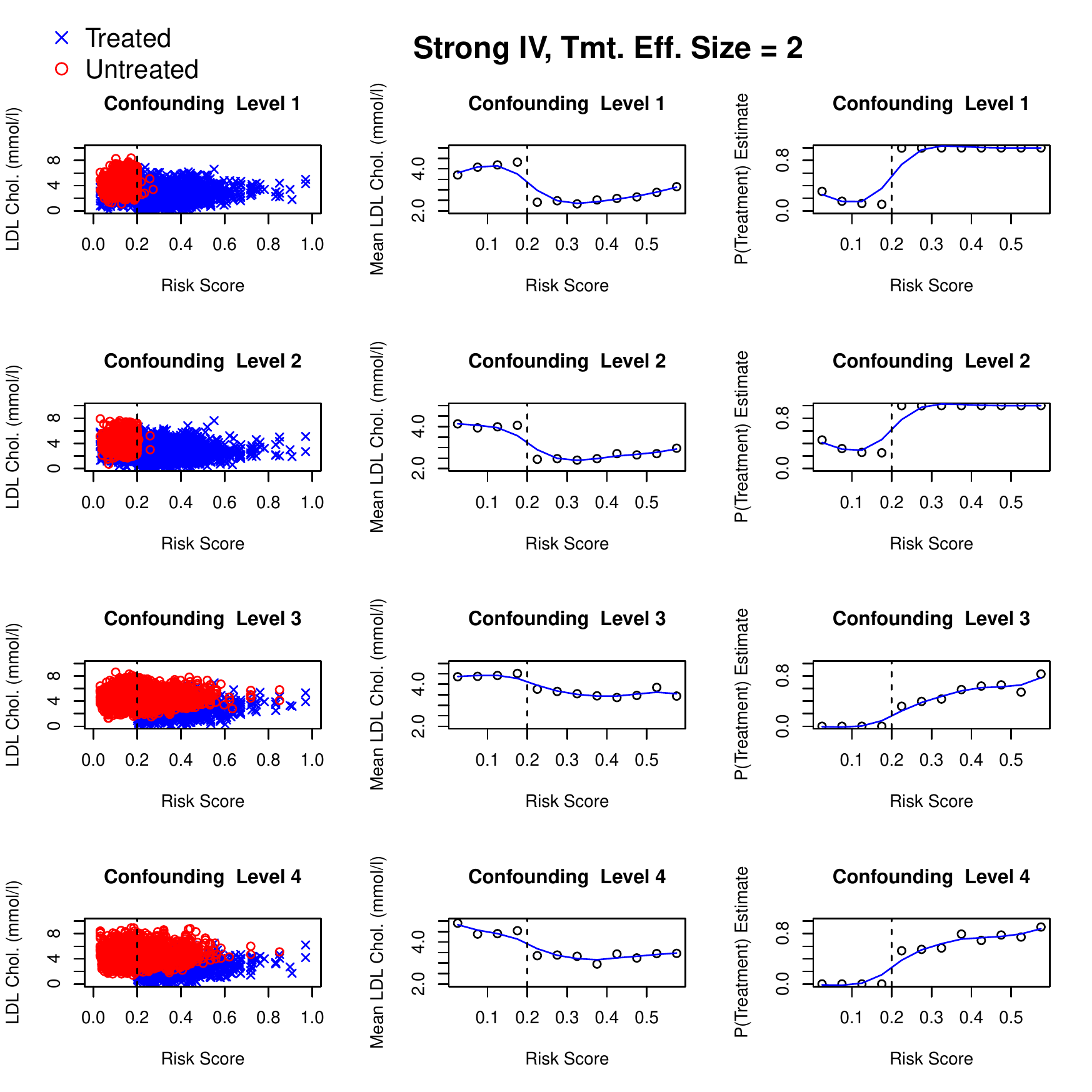}
\end{center}
\caption{\label{fig:usefulplots_strong}Plots in the left hand column show risk vs. simulated LDL cholesterol level, those in the central column show risk score (bin mid-point) vs. sample mean LDL cholesterol level and those in the right-hand column show risk score (bin-midpoint) vs. estimated probability of treatment. Plots are shown for different levels of confounding using simulated datasets with a treatment effect of size 2 and threshold acting as a strong instrument for treatment. A dashed vertical line indicates the threshold level.}\label{fig1}
\end{figure}

\begin{figure}[!h]
\begin{center}
\includegraphics[scale=0.7]{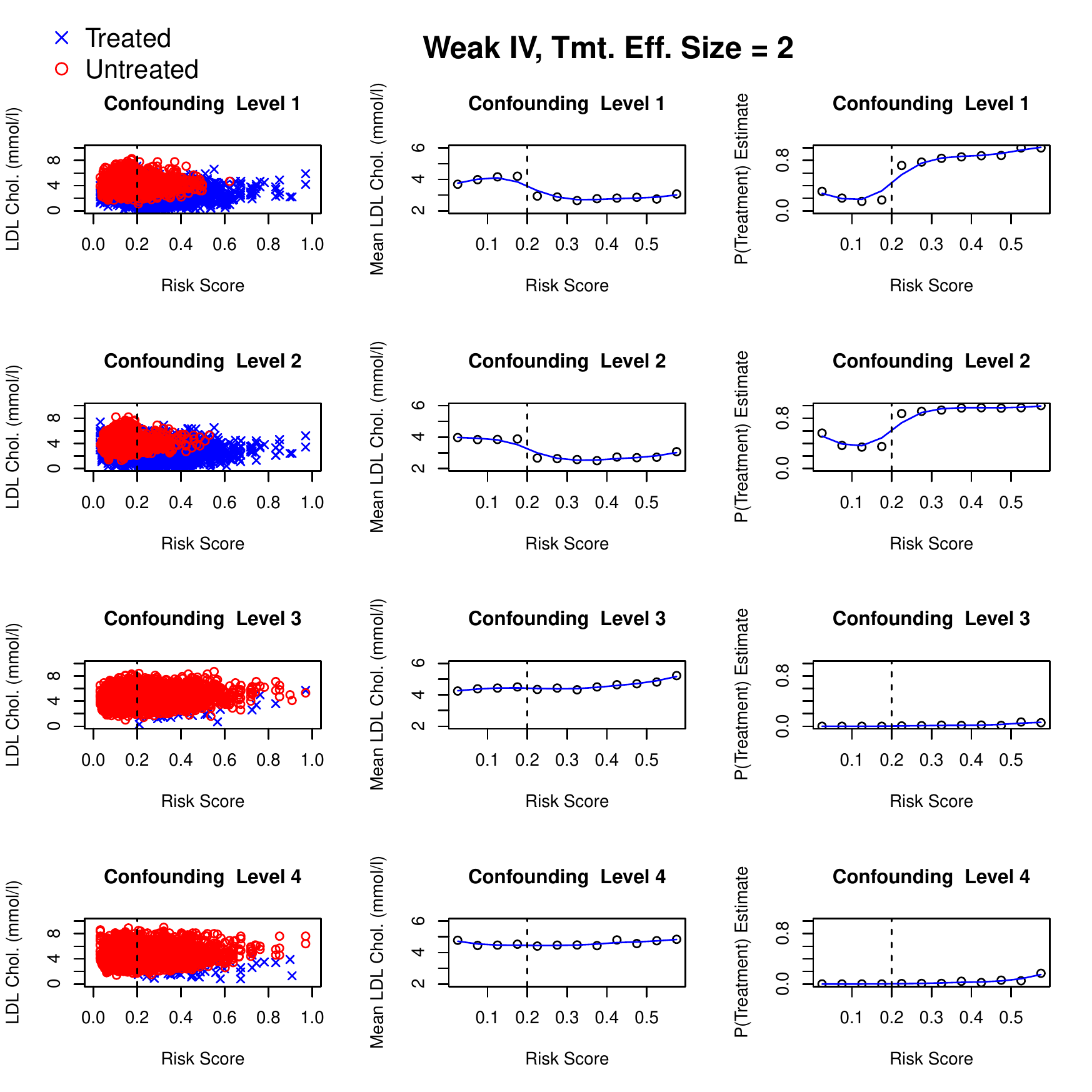}
\end{center}
\caption{\label{fig:usefulplots_weak}Plots in the left hand column show risk vs. simulated LDL cholesterol level, those in the central column show risk score (bin mid-point) vs. sample mean LDL cholesterol level and those in the right-hand column show risk score (bin-midpoint) vs. estimated probability of treatment. Plots are shown for different levels of confounding using simulated datasets with a treatment effect of size 2 and threshold acting as a weak instrument for treatment. A dashed vertical line indicates the threshold level.}\label{fig2}
\end{figure}

We simulated 100 datasets using the algorithm described in Section \ref{sec:sims} and fitted models using each of them. It is often not clear whether or not there exists a discontinuity at particular threshold, especially when data are very variable. We investigated this further by producing plots of the raw data points (continuous threshold variable against outcome) and by producing plots showing outcome mean estimate and raw probability of treatment estimate within regular bin widths defined by the threshold variable (in this case, the risk score). This is a common initial exploratory analysis when an RD design is thought to be appropriate and is typically used as a tool to back up the assumptions which determine whether or not an RD design is valid \cite{ImbensLemieux2008, Lee2008, Leeetal2004, Lalive:08}. Figures \ref{fig:usefulplots_strong} and \ref{fig:usefulplots_weak} show such plots produced using one of the simulated datasets described above, under each defined level of unobserved confounding and instrument strength for the design threshold. In each case the treatment effect size is 2. The plots in Figure \ref{fig:usefulplots_strong} were produced using datasets in which threshold is a strong instrument for treatment, whereas Figure \ref{fig:usefulplots_weak} was produced using datasets where threshold is a weak instrument for treatment. The raw plots (left-hand column) show clearly that the RD design becomes more fuzzy as confounding increases, especially where the threshold is a weak instrument for treatment. The plots of the mean outcomes (central column) and estimated probabilities of treatment (right-hand column) show obvious discontinuities at the threshold value of 0.2, except in Figure \ref{fig:usefulplots_weak} where the level of unobserved confounding is high (levels 3 or 4). The discontinuities are generally larger for lower levels of unobserved confounding. When plots of either the estimated outcome means or raw estimates of probability of treatment -- within risk score bins -- exhibit a jump in at the threshold, then there is some evidence to suggest the use of the RD design may be appropriate. In light of these initial plots, an attempt to implement the RD design appears reasonable in all scenarios except where threshold is a weak instrument for treatment and unobserved confounding is thought to be at a high level. We performed analyses using RD designs on each of the 100 simulated datasets for all levels of unobserved confounding and instrument strengths for threshold. Results were combined for each unobserved confounding/instrument strength level and we now present some of these results. 

On performing the analyses using the RD design models, we found that, across all considered bandwidths and treatment effects, data at confounding levels 1 and 2 showed similar results and, in addition, data at confounding levels 3 and 4 showed similar results for both instrument strengths. This is perhaps not too surprising since the only difference between these scenarios is in the estimated correlation between LDL cholesterol level and HDL cholesterol level. Hence, for brevity, we present tables of results that only include unobserved confounding levels 1 (low level of unobserved confounding) and 3 (high level of unobserved confounding). Furthermore, we show results for a simulated treatment effect of size 2 and for chosen bandwidths 0.05 and 0.25. A bandwith of 0.15 was also considered in addition to treatment effect sizes of 0.5 and 1.09, across all three bandwidths. All results are available in full within the online supplementary material at the webpage \url{www.statistica.it/gianluca/RDD}.

Tables~\ref{results_tab1} and \ref{results_tab2} show results (treatment effect estimates together with associated 95\% confidence or credible intervals) from the simulation studies with treatment effect set to 2 (\ie treatment with statins is associated with a reduction of 2mmol/l) for chosen bandwidths 0.05 and 0.25, respectively. We include results using ATE estimators obtained by estimating the regression model \bref{eqn:sharpest} using a standard frequentist analysis, which we term $\Delta_\beta^{\rm{\textit{freq}}}$, along with all Bayesian estimators described in Section \ref{sec:bayes}

\begin{table}[!h]
\begin{center}
\scalebox{0.8}{
\begin{tabular}{c c c cc ccc}
\multicolumn{7}{l}{\em Bandwidth = 0.05, Treatment Effect Size = 2}\\
\hline
~ & ~ & \multicolumn{3}{c}{ATE Estimators} & \multicolumn{3}{c}{LATE Estimators} \\
\em{IV} & \em{Confounding}  & $\Delta^{\rm{\it{freq}}}_\beta$ & $\Delta^{\rm{\it{wip}}}_\beta$ & $\Delta^{\rm{\it{sip}}}_\beta$ & $\mbox{LATE}_{\rm{\it{unct}}}$ & $\mbox{LATE}_{\rm{\it{flex}}}$ & $\mbox{LATE}_{\rm{\it{cnst}}}$\\
\hline
Strong & 1: LOW & -1.74 & -1.86 & -1.87 & -2.10 & -2.10 & -2.10 \\
~ & ~ & (-1.98, -1.51) & (-1.98, -1.74) & (-1.99, -1.74) & (-2.25, -1.95) & (-2.26, -1.96) & (-2.24, -1.95) \\
~ & 3: HIGH & -0.74 & -0.89 & -0.90 & -2.20 & -2.20 & -1.75 \\
~ & ~ & (-1.08, -0.41) & (-1.02, -0.76) & (-1.03, -0.76) & (-2.59, -1.83) & (-2.59, -1.83) & (-2.03, -1.48) \\
\hline
Weak & 1: LOW & -1.01 & -1.16 & -1.17 & -2.19 & -2.18 & -1.84 \\
~ & ~ & (-1.31, -0.72) & (-1.29, -1.03) & (-1.30, -1.04) & (-2.49, -1.91) & (-2.48, -1.90) & (-2.07, -1.62) \\
~ & 3: HIGH & 0.05 & -0.08 &  -0.09 & -45.72 & -15.75 &  -0.51 \\
~ & ~ & (-0.16, 0.25) & (  -0.20,   0.04) & (  -0.21,   0.03) & (-311.52, 207.84) & ( -87.39,  29.38) & (  -1.23,   0.20) \\
\hline
\end{tabular}
}
\caption{\label{results_tab1} Simulation study results over 100 simulated datasets, for various confounding scenarios and instrument strengths for threshold. Intervals are 95\% credible intervals or, for non-Bayesian estimates, 95\% confidence intervals. Treatment effect size = 2, bandwidth = 0.05.}
\end{center}
\end{table}

\begin{table}[!h]
\begin{center}
\scalebox{0.8}{
\begin{tabular}{c c c cc ccc}
\multicolumn{7}{l}{\em Bandwidth = 0.25, Treatment Effect Size = 2}\\
\hline
~ & ~ & \multicolumn{3}{c}{ATE Estimators} & \multicolumn{3}{c}{LATE Estimators} \\
\em{IV} & \em{Confounding} & $\Delta^{\rm{\it{freq}}}_\beta$ & $\Delta^{\rm{\it{wip}}}_\beta$ & $\Delta^{\rm{\it{sip}}}_\beta$ & $\mbox{LATE}_{\rm{\it{unct}}}$ & $\mbox{LATE}_{\rm{\it{flex}}}$ & $\mbox{LATE}_{\rm{\it{cnst}}}$\\
\hline
Strong & 1: LOW & -2.02 & -1.98 & -1.98 & -2.26 & -2.26 & -2.26 \\
~ & ~ & (-2.17, -1.87) & (-2.08, -1.88) & (-2.08, -1.89) & (-2.38, -2.14) & (-2.38, -2.14) & (-2.37, -2.14) \\
~ & 3: HIGH & -0.97 & -0.94 & -0.94 & -1.90 & -1.90 & -1.78 \\
~ & ~ & (-1.27, -0.67) & (-1.04, -0.83) & (-1.05, -0.84) & (-2.14, -1.66) & (-2.14, -1.66) & (-1.99, -1.56) \\
\hline
Weak & 1: LOW & -1.25 & -1.24 & -1.25 & -2.11 & -2.10 & -1.92 \\
~ & ~ & (-1.47, -1.04) & (-1.35, -1.14) & (-1.35, -1.14) & (-2.31, -1.91) & (-2.31, -1.91) & (-2.09, -1.75) \\
~ & 3: HIGH & -0.20 & -0.18 &  -0.19 & -25.28 & -22.85 &  -2.51 \\
 ~ & ~ & (-0.31, -0.08) & ( -0.27,  -0.08) & ( -0.28,  -0.09) & (-49.48, -10.15) & (-48.68,  -9.12) & ( -3.88,  -1.22) \\
\hline
\end{tabular}
}
\caption{\label{results_tab2} Simulation study results over 100 simulated datasets, for various confounding scenarios and instrument strengths for threshold. Intervals are 95\% credible intervals or, for non-Bayesian estimates, 95\% confidence intervals. Treatment effect size = 2, bandwith = 0.25.}
\end{center}
\end{table}

 
Examining Tables~\ref{results_tab1} and \ref{results_tab2}, we see that the Bayesian LATE estimators generally capture the true value of the treatment effect (-2.00) and provide plausible 95\% credible intervals for both confounding levels where threshold is a strong instrument for treatment and for the low unobserved confounding level where threshold is a weak instrument for treatment. In general, both Bayesian and non-Bayesian ATE estimators do not tend to reflect the true treatment effect, especially as the unobserved confounding level increases and the strength of threshold as an instrument weakens. An exception is when the bandwith is large (0.25), the level of unobserved confounding is low and the threshold is a strong instrument for treatment. This may be expected as the RD design might be considered almost sharp where threshold is a particularly strong instrument for treatment. In addition, a relatively large bandwidth of 0.25 ensures that there are many treated individuals above the threshold and many untreated individuals below the threshold and, in such cases, an ATE estimator may be considered appropriate. The larger amount of utilised data for the bandwidth of 0.25 may also explain why the frequentist ATE estimates are more similar to the Bayesian ATE estimates in Table~\ref{results_tab2} when compared to those in Table~\ref{results_tab1}. In general, there is some bias in most estimates, possibly as a result of different sources of noise incorporated into the simulation set-up, together with unobserved confounding and changing instrument strength.

Where unobserved confounding is high and the threshold is a weak instrument for treatment, we see that all estimators behave in an unpredictable manner and fail to estimate the treatment effect accurately. This is not surprising, since the plots presented in Figures \ref{fig:usefulplots_strong} and \ref{fig:usefulplots_weak} implied that an RD design was not appropriate for these scenarios, since the design becomes too fuzzy for the modelling techniques presented to be applicable. Similar problems are seen in simulation studies investigating the effect of weak instruments with unobserved confounding \cite{Jonesetal2012}.

\subsection{Sensitivity to prior specification} 
\label{sec:priorspec}
We considered a number of prior specifications in this work. In situations where such information was available, for example the possible size and nature of the effect of statins on LDL cholesterol levels based on clinical trial results and/or expert GP knowledge, we attempted to account for this. Where less information was available, as in the case of the probabilities in the denominator for the LATE, we attempted to understand the sensitivity of results to prior specification. 

Overall, the effect of the prior information appears to be negligible for the ATE, with the $\Delta^{\rm{\it{wip}}}_\beta$ and $\Delta^{\rm{\it{sip}}}_\beta$ ATE estimators producing similar estimates across all scenarios and for both bandwidths. Similarly, there are no obvious differences between the $\mbox{LATE}_{\rm{\it{unct}}}$ and $\mbox{LATE}_{\rm{\it{flex}}}$ estimators under these different prior distributional assumptions. The $\mbox{LATE}_{\rm{\it{cnst}}}$ estimator yields a slightly higher estimate than both the $\mbox{LATE}_{\rm{\it{unct}}}$ and $\mbox{LATE}_{\rm{\it{flex}}}$ where there is a high level of unobserved confounding (Level 3) and threshold is a strong instrument for treatment and also where there is a low level of unobserved confounding and threshold is a weak instrument for treatment. This is due possibly to the size of the denominator in this estimator, which is always slightly bigger than those seen in the other two Bayesian LATE estimators. We fixed the minimum size of denominator by imposing at least 200 more treated above than below the threshold. Thus, even if the denominator is typically greater that this minimum, on average over all the simulations, this will result in a larger LATE. This difference is more pronounced in the case of the smaller 0.05 bandwidth than where a bandwidth of size 0.25 was used. This feature of the $\mbox{LATE}_{\rm{\it{cnst}}}$ may also explain why this estimator produced a reasonable estimation of the treatment effect where unobserved confounding is high and threshold is a weak instrument for treatment, using a bandwidth of 0.25 only (Table~\ref{results_tab2}). Overall, flexible priors seem to provide the most reliable results for LATE estimators in situations where, based on the exploratory plots, an RD design is deemed appropriate. In these cases when a treatment effect is present, the effect is recovered even when the IV is weak or, in the case of a strong IV, when the level of unobserved confounding is high. 

\section{Discussion}
\label{sec:discussion}
\subsection{Critical issues}
\subsubsection{``Local'' vs ``global'' effect}
An apparent drawback of the RD design is the ``local'' nature of the causal estimate, \ie there is no guarantee that the causal effect is the same over the whole range of the risk score. If the aim of estimating the causal effect is to compare it to the results of trials and to determine whether the prescription guidelines are effective, the local nature is not a disadvantage. Rather it will highlight whether the guidelines need to change if the results are starkly different from those of a (well conducted) trial. Furthermore, while trials may indicate that the effect of statins is constant across strata of age, sex and initial cholesterol levels, there is no reason to assume that this applies across risk scores in the general population treated by GPs, especially when partial compliance of patients to prescriptions is to be expected. In Section \ref{sec:future} below we discuss how multiple thresholds might be used to determine whether the effect is constant across the range of the assignment variable.

\subsubsection{Compliance and adherence}
In the context of the case study on which our simulations are based, we have two types of ``compliance''. One is the adherence of the GP to the prescription guidelines, which we have assumed to be partial, in our simulations. The second is the compliance of the patient to the treatment prescription, which in contrast we have assumed is perfect. In real data, this is hardly ever the case: many patients do not take statins when they have been prescribed. 

This aspect also relates to the fact that the LATE estimates a causal
effect of a treatment in a population defined by the fact that the GP
adhered to the prescription guidelines. We can ask two questions
here. First; are patients whose GPs adhere to
guidelines comparable to those whose GPs have alternative
strategies? Second; given that we are interested in comparing the RD design results
from primary care to those of RCTs, are RCT
participants comparable to patients whose GPs adhere to
guidelines? 

The first question means we need to understand whether GPs who
prescribe according to the guidelines have patients that are
systematically different from those who have GPs with alternative
treatment strategies. There might be circumstances where this is
the case, e.g. if different primary care trusts have different
treatment ``cultures'' as well as different patient
populations. Another context where this might be the case is if
`strictly adherent' GPs prescribe only according to guidelines whilst `non-adherent' GPs do not
and they serve different populations. It is hard to believe either of
these would be true, especially given the fact that the guidelines are
nationwide and any individual quirks of GPs would probably even out over the
population. Thus, we are reasonably confident that the LATE is informative about the
general primary care population.

In answer to the second question, we must consider that individuals
recruited into an RCT are often selected on the basis of
characteristics that make them more likely to comply and that a
primary care population will not necessarily be similar in those
respects. Thus, we might
expect that if we do not take into account the probable lack of
patient compliance, we would see a smaller effect size for statins in
primary care than in RCTs. In order to deal with this in future work,
we will focus on subgroups of primary care patients who can be
considered in some way exchangeable with RCT recruits.


\subsection{Future work}
\label{sec:future}
This leads us into the potential problem with Assumption 5 which is necessary to identify the LATE. This
assumption states that there are no GPs whose prescription strategy
is to refuse to adhere to the guidelines. This only makes sense if we
believe that GPs have treatment strategies in place before seeing
patients and that they act according to these strategies. While this
seems plausible when referring to GPs, this is not always the case
when applied to patient compliance. In this case, we would be
requiring that patients have strategies regarding compliance to taking
medication in place before they are prescribed and that they act in
accordance to these strategies. Moreover, that there are no
patients whose strategy it is to ``defy'' the prescription. Both
aspects of the assumption are less credible as patients are less
likely to have strategies and there are likely to be patients who will
try to do the opposite of what they are ``told''. We mention this here
in order to support our use of the LATE and to distinguish it from the
more common situation of patient compliance where it is used and
potentially less reliable. In dealing with patient compliance we recommend limiting the RD design to those patients whom we consider exchangeable, so that we may not need to introduce additional complexity within the models to account for patient non-compliance. Further work in this respect is required but is outside the scope of this paper.

Our focus has been here on statin prescription, where strong
information can be brought to bear in prior model formulation. With
other treatments and outcomes, it may be that there is limited
knowledge regarding the effect of the treatment on the outcome
(generally to a specific sub-population of patients) or of
clinical adherence to treatment guidelines, but that there exists a
vast amount of real observational data in primary care. We believe that
it would be useful to apply Bayesian RD methods in such a
scenario to combine limited evidence-based and clinical prior beliefs
with actual observed data in an effort to assess treatment effects in
clinical practice and perhaps inform whether or not further
trials/experiments should be considered. 


We believe that the RD design has a great potential in primary care. We can imagine
that in the future, trial results will be augmented by planned RD
designs with thresholds at different levels of the assignment variable
in order to determine where in disease progression the treatment is
most effective in primary care as well as having a more realistic
basis for cost-effectiveness analyses. This is particularly relevant
when the treatment targets individuals who are likely to be extreme
and under-represented in trials, or when the treatment is for specific
subgroups of the population, such as terminally ill patients. Additional model assumptions or adjustments may be required when fitting an RD design to such subgroups.

\section{Acknowledgments}
This research has been funded by a UK MRC grant MR/K014838/1. We wish to thank Prof Nick Freemantle, Dr Irene Petersen, Prof Richard Morris, Prof Irwin Nazareth and Prof Philip Dawid for providing insightful and thought-provoking comments.

\bibliography{rdd_arXiv}
\bibliographystyle{wileyj}

\newpage

\end{document}